# Accelerated Increase in the Decay of Radioactive $^{125}$I by X- Ray Irradiation

By

S. Soloway (now retired)

*Experiments done at Brookhaven National Laboratory*

**Expanded concepts of photonuclear reaction in Mossbauer type behavior were applied to radioactive nuclei. An enhanced reduction in radioactivity of $^{125}$I was achieved by X-ray irradiation using the Brookhaven Synchrotron.**

The interaction of photons with selected stable nuclei has been an active field in Mossbauer research for decades. Resonant photons excite a stable nucleus to a higher energy level following which the nucleus emits the absorbed photon and drops to its original ground state. Properties in the Mossbauer studies are unique and may be represented by the popular $^{57}$Fe. The resonant photon for $^{57}$Fe is 14.4 keV. It is generated by the 270 day decay of $^{57}$Co generating a 14.4 keV photon during its decay to the ground state of $^{57}$Fe. The emitted photon has the incredibly narrow bandwidth of $10^{-5}$ ev. The capture cross section for this photon by the $^{57}$Fe nucleus is greater than $10^{6}$ barns, again enormous. The excited $^{57}$Fe nucleus drops to its ground state in $10^{-7}$ to $10^{-9}$ seconds.

The project reported here is: What happens if the Mossbauer concept is applied to a radioactive nucleus? We could anticipate that a radioactive nucleus could be excited to a higher energy following which it would emit the absorbed photon in say $10^{-7}$ seconds and drop back to the original radioactive nucleus. Its radioactivity properties would be essentially unchanged.

Radioactive nuceli are, of course, different from stable nuclei. As a result of photon absorption, the radioactive properties may be changed; its half-life could be increased or decreased. Possibly were the process of photon absorption to mimic "stimulated emission" as expressed by lasers, there could be a major speed up in radioactive decay.

To perform this kind of exploratory experiment, a unique Mossbauer spectrophotometer would be desirable. Its photon energy range would be 1kHz to 8 MeV. Photon bandwidth would be $10^{-5}$ ev. And photon flux should be very large. Such a spectrophotometer is, of course, unavailable.

Certain compromises are possible in a feasibility experiment. A Synchrotron with its large photon flux would be possible. Even though the photon bandwidth is broad compared with $10^{-5}$ ev, together with a small flux in that bandwidth, the photon flux could be maintained for weeks. And the effectiveness of the photon exposure of a radioactive nucleus would be determined by comparing a photon exposed nucleus with the normal decay of a control radioactive nucleus.

The X-19B beamline of the Brookhaven Synchrotron was used in the exploratory experiments described below. Figure 1 shows Brightness versus Photon Energy for this beam line. At this time, we do not know the precise resonant photon excitation spectra for radioactive nuclei. A broad energy spectrum is needed for a feasibility study—say from 10 ev to 30,000 ev as represented by Figure 1. For maximum photon flux, the radioactive samples should be as close to the beam source as possible, here 3.5 meters. Peak flux occurs at 5000 ev. The 0.1% bandwidth results in 5 ev for an output flux of about $2 \times 10^{13}$ photons/sec/0.1% bandwidth/mrad$^2$. This leads to $0.4 \times 10^8$ photons/sec/mrad$^2$ for a bandwidth of $10^{-5}$ ev. This flux decreases by a factor of 50 at the extremes of the photon energy range.

The possible radioactive isotopes that could be chosen are, of course, very large. The one chosen was based on safety in the conduct of the experiment, namely minimizing the possible contamination of the Synchrotron. Encapsulated $^{125}$I source was chosen, manufactured by Amersham Health Care [1] and widely used for tumor irradiation. $^{125}$I decays by K or L capture to $^{125}$Te with a 60 day half-life which then emits a 35.4 keV gamma ray in dropping to its ground state in 1.49 x $10^{-9}$ seconds. The seed is shown in Figure 2. The enclosure is a sealed titanium container stable to $600^0$C.

$^{125}$I is chemically bound to Ag in the seed and AgI is stable to $400^0$C. Clinically the seeds are routinely sterilized to $120^0$C to $130^0$C. Seed activity that was used, dictated by vendor company, was about 0.05 mc.

Before the experiment was initiated, the beam, where seeds were to be placed, was located by film exposure, not very precise but adequate.

To determine whether beam heating of the seed were to occur, leading to possible seed rupture, the temperature at the peak beam location was measured. Temperature rise was trivial and Radiation Safety Analysts considered that no problem existed.

For the precise measurement of the seeds before and after photon beam exposure, an Auto Gamma Series Packard counter [2] was made available. It is routinely used in clinical laboratories. A NaI crystal was set with a window from 15 keV to 80 keV. As such, it detected 35.4 keV of $^{125}$Te as well as K and L capture X-rays as well as X-rays from Ag fluorescence. Repeated counting measurements of a seed gave a repeatability of 0.5% with counting rates of the seeds 0.1 to 1.0 x $10^6$ cpm.

Figure 3 shows the organization of four seeds exposed to the Synchrotron beam in Experiment 1 where seed #1 was exposed to the maximum beam intensity as determined by film discoloration earlier. In addition two control seeds were located at an office remote from the Synchrotron. Diverse experiments carried out at other beams of the Synchrotron during the exposure period for this first experiment determined the available exposure time in the first experiment to 930 hours, about 39 days.

The results of Experiment 1 are shown in Table 1. Counting rates of the seeds at the beginning of the experiment as well as counting rate 930 hours later are recorded. The ratio of final to initial counting is recorded for each seed. The ratio represents, for controls, the fraction of the original radioactivity remaining after 930 hours of normal decay.  Seed #1 exposed to the photon beam beam has lost an excess of 12.38% more radioactive nuclei; seed #2 has lost 2.00% more than the controls.

A repeat experiment was run with the orientation in Figure 4 with an increased number of seeds.  Exposure of seeds in photon beam this time was 1129 hours (47 days), exposure time again dictated by the Synchrotron schedule.  Seed #7 shows a reduced radioactivity from the controls of –13.87%, seed #8 was –6.80%; seed #9 was –3.87%; seed #12 was –3.36%.  This confirms the results of Experiment 1.

The following evaluates some considerations as to whether the reduced level of radioactivity could be due to factors other than photon exposure.

The electron storage ring of the Synchrotron could generate sources of radiation other than the photon beam shown in Figure 1.  Bremstrahling radiation could be generated from electrons impinging residual gas in the ring chamber with an endpoint energy of 2.584 GeV.  Also these Bremstrahling gamma rays could create neutrons by giant dipole resonance reaction with residual gas.  Such neutrons and/or Bremstrahling gammas cooould affect the $^{125}$I level of radioactivity.  However, in both cases, should this occur, all seeds would have been illuminated uniformly.  This effect does not seem to be occurring.

A second possible effect could arise from the large Synchrotron photon flux causing the rupture of the AgI compound leading to the $^{125}$I migrating inward into Ag or outward through a possible hole in the titanium container.
AgI as noted earlier is stable to $400^0$C and is constantly being bombarded with 35.4 keV $^{125}$Te gamma rays from $^{125}$I decay together with the accompanying K and L X-rays and the resulting Ag fluorescence.  This constant and large photon exposure has never been a factor in the normal use of these seeds for radiation exposure in therapy for decades.  As to a rupture of the seed, the enclosure was designed and evaluated up to $600^0$C.  In addition, the Health Physics Personnel at Brookhaven did extensive wipe tests after each experiment with a sensitivity of 10 picocuries. This chemical rupture and movement of $^{125}$I does not seem to be occurring.

The role of photons interacting with radioactive nuclei will require more research to understand the actual mechanism in the reaction reported here.

The results detailed here were reported orally at an American Nuclear conference [3]. Since that time, limited reports have appeared involving radioactive nuclei and photons [4] and [5]. In addition, an excellent summary has appeared "Energy Traps in Nuclei"[6] discussing the possible mechanisms for the interaction of photons with the nucleus.

I want to acknowledge the extensive cooperation and knowledgeable help provided me by Dr. James P. Harbison and Dr. Scott M. Soloway.[1]Amersham Health Care Medi-Physics, Inc.

  2636 S. Clearbrook Drive

  Arlington Heights IL 60005

[2]Packard Instrument Company

  800 Research Parkway

  Meriden, CT 06450

[3]S. Soloway and J. Harbison, Am. Nuclear Society Transactions, Biol. and

  Medicine, 75,20, 1996

[4]C.B. Collins et al, Phys. Rev. Letters 82, No. 4, (pp.695-698), 25 Jan 1999

[5]C. Platt, Wired, p.72,Jan 1999

[6]P. Walker and G. Dracoulis, Nature 399 (pp.35-340) 6 May,19995

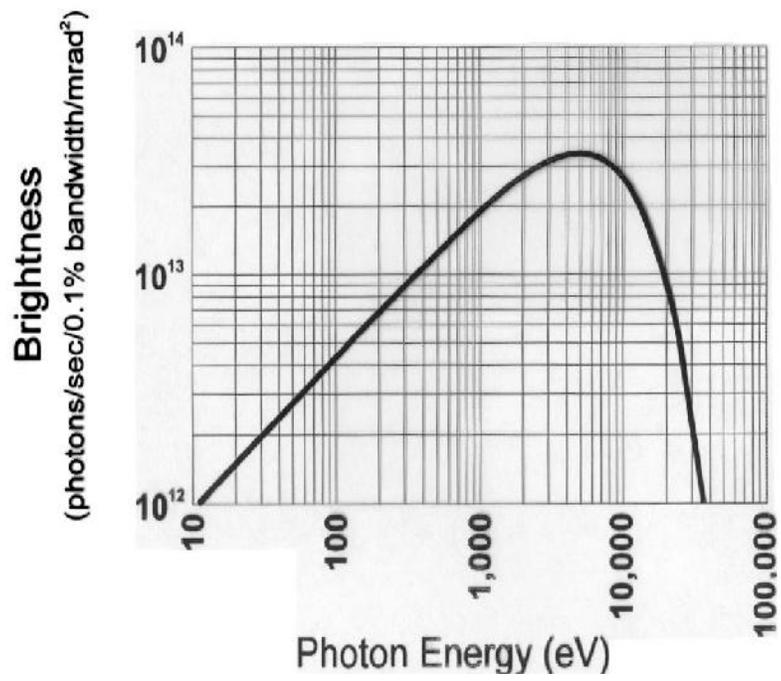

**Figure 1**

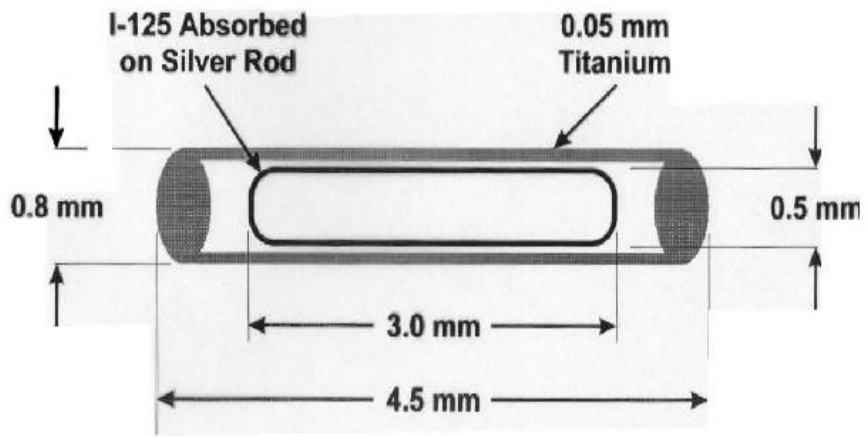

Figure 2

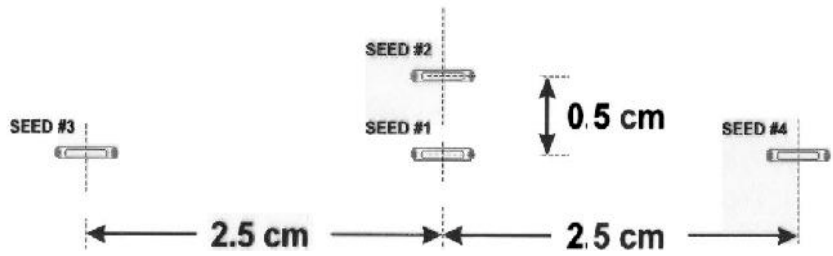

**Figure 3**

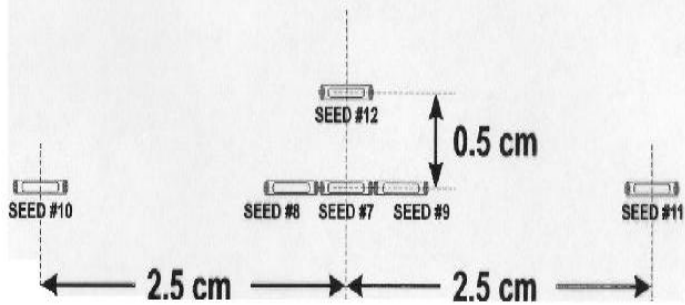

**Figure 4**

TABLE 1

EXPERIMENT 1

| Seed | Seed Location | Initial Activity X $10^6$ cpm | Final Activity (930 hrs) x $10^6$ cpm | Ratio | % from Expected Radioactivity |
|---|---|---|---|---|---|
| 1 | Main Beam | 1.756 | 1.038 | 0.591 | -12.38% |
| 2 | 0.5 above | 1.700 | 1.123 | 0.661 | -2.00% |
| 3 | 2.5 cm left | 1.814 | 1.219 | 0.672 | -0.52% |
| 4 | 2.5 cm right | 1.807 | 1.212 | 0.671 | -0.52% |
| 5 | Control | 1.888 | 1.273 | 0.674 | -0.01% |
| 6 | Control | 1.838 | 1.240 | 0.675 | +0.01% |

# TABLE 2

EXPERIMENT 2

| Seed | Seed Location | Initial Activity x $10^6$ cpm | Final Activity (1129 hrs) x $10^6$ cpm | Ratio | % from Expected Radioactivity |
|---|---|---|---|---|---|
| 5 | Control | 1.258 | 0.733 | 0.583 | +0.43% |
| 6 | Control | 1.228 | 0.710 | 0.578 | -0.43% |
| 7 | MAIN BEAM | 1.195 | 0.598 | 0.500 | -13.86% |
| 8 | SLIGHT LEFT | 1.187 | 0.642 | 0.541 | -6.80% |
| 9 | SLIGHT RIGHT | 1.208 | 0.674 | 0.558 | -3.87% |
| 10 | 2.5 CM RIGHT | 1.239 | 0.719 | 0.581 | +0.09% |
| 11 | 2.5 CM LEFT | 1.226 | 0.715 | 0.583 | +0.43% |
| 12 | 0.5 CM ABOVE | 1.239 | 0.695 | 0.561 | -3.36% |